\documentclass[useAMS,usenatbib]{mn2e} 
\usepackage{graphicx,amssymb}

\title[Phase-Space Density of Dark Matter Haloes]{Is the Radial Profile of the 
Phase-Space Density of Dark Matter Haloes a Power-Law?}
\author[Ma, Chang, \& Zhang]{Chung-Pei Ma$^{1}$\thanks{E-mail: cpma@berkeley.edu (CPM), pchang@astro.berkeley.edu (PC), jzhang@astro.berkeley.edu (JZ)}, Philip Chang$^{1}$, and Jun Zhang$^{1}$\\
$^{1}$Department of Astronomy and Theoretical Astrophysics Center, 601 Campbell Hall, University of California, Berkeley, CA 94720
}

\pagerange{ 
\pageref{firstpage}-- 
\pageref{lastpage}}

\usepackage{multirow}
\usepackage{subfigure}

\def \rsig {\rho/\sigma^3_r}

\usepackage{amsmath} 
\usepackage{multirow}

\begin{document}

\label{firstpage}

\maketitle 
\begin{abstract}
  The latest cosmological $N$-body simulations find two intriguing
  properties for dark matter haloes: (1) their radial density profile,
  $\rho$, is better fit by a form that flattens to a constant at the halo
  center (the Einasto profile) than the widely-used NFW form; (2) the
  radial profile of the pseudo-phase-space density, $\rsig$, on the other
  hand, continues to be well fit by a power law, as seen in earlier
  lower-resolution simulations.  In this paper we use the Jeans equation to
  argue that (1) and (2) cannot both be true at all radii.  We examine the
  {\it implied} radial dependence of $\rsig$ over 12 orders of magnitude in
  radius by solving the Jeans equation for a broad range of input $\rho$
  and velocity anisotropy $\beta$.  Independent of $\beta$, we find that
  $\rsig$ is approximately a power law {\it only} over the limited range of
  halo radius resolvable by current simulations (down to $\sim 0.1$\% of
  the virial radius), and $\rsig$ deviates significantly from a power-law
  below this scale for both the Einasto and NFW $\rho$.  The same
  conclusion also applies to a more general density-velocity relation
  $\rho/\sigma_D^\epsilon$.  Conversely, when we enforce $\rsig \propto
  r^{-\eta}$ as an input, none of the physically allowed $\rho$ (occurring
  for the narrow range $1.8 \la \eta \le 1.9444$) follows the Einasto form.
  We expect the next-generation simulations with better spatial resolution
  to settle the debate: either the Einasto profile will continue to hold
  and $\rsig$ will deviate from a power law, or $\rsig$ will continue as a
  power law and $\rho$ will deviate from its current parameterizations.
\end{abstract}

\section{Introduction} \label{introduction}

$N$-body simulations of cosmological structure formation have shown that
the spherically-averaged radial profiles of the mass density and velocity
dispersion of dark matter haloes follow simple and nearly universal
functional forms that are largely independent of halo properties such as
mass, environment, and formation history.  For the density profile $\rho$,
there had been considerable discussion about the value of the logarithmic
slope of its central cusp, $\gamma \equiv d\log \rho/d\log r$, whether it
is $-1$ as in the forms of, e.g., \citet{Hernquist90, NFW97}, or $-1.5$ as
in \citet{Moore99}.  Results from recent $N$-body simulations now suggest a
lack of a definite inner slope -- the density profile of these
better-resolved dark matter haloes continues to flatten with shrinking
radius (e.g., \citealt{Navarro04, Merritt05, Merritt06, Graham06,
  Navarro08, Stadel08}).  Functional forms such as \citet{Einasto69} and
\citet{PS97} motivated by the Sersic profile for the surface brightness of
galaxies \citep{Sersic68} appear to provide a more accurate fit to the
latest simulations.  In these forms, $\gamma$ itself is a function of
radius and asymptotes to zero at the halo center.
%The former (in 3-d) has the same form as the
%Sersic profile (cite) commonly used to fit the 2-d projected surface
%brightness of galaxies, while the latter is the de-projected 3-d profile of
%the Sersic form.

A second property that has attracted much attention lately is the
radial profile of the pseudo-phase-space density, $\rho/\sigma^3$, which
has been reported to be well approximated by a power-law in a number of
$N$-body simulations (e.g., \citealt{TN01, Ascasibar04, DM05, Hoffman07,
  Stadel08, Navarro08}).  Both the total velocity dispersion and the
velocity dispersion in the radial direction have been used to define
$\sigma$.  Unlike the controversial density profile, whose best-fit form
has changed over the years with improved numerical resolution, the
power-law profile of $\rho/\sigma^3$ has withstood the scrutiny, and
different studies have all reported similar findings except for a minor
variation in the actual value of the slope of the power law.

A third relation was proposed when the velocity anisotropy,
$\beta(r)=1-\sigma_t^2/\sigma_r^2$, of simulated haloes was taken into
account.  \citet{HM06} advocated a linear relation between $\beta$ and the
local logarithmic slope of $\rho$.  Other studies, however, have found a
large scatter in this relation, particularly in the outer parts of the
haloes beyond the scale radius \citep{DM05, Navarro08}.

To help elucidate the physical meanings of these empirical relations
determined from simulated haloes, a typical approach is to use the Jeans
equation for a spherical, self-gravitating collisionless system in
equilibrium to predict the density or velocity structures of
dark matter haloes under a certain set of assumptions.  Most of the recent
studies based on this approach have begun with the assumption of a
power-law $\rsig(r)$.  These papers then explored the density profiles
allowed by the Jeans equation with either isotropic velocities
(e.g. \citealt{TN01,Hansen04}), or an anisotropic velocity profile
$\beta(r)$ (e.g., \citealt{DM05}; Hansen Stadel 2006).  Alternatively, some
authors have chosen to study the radial profile of the velocity
anisotropy $\beta(r)$ from the Jeans equation, starting with a power-law
$\rsig$ and some input form for $\rho$ (e.g., \citealt{Zait08}).

In this paper, we reexamine this question by taking a complementary
approach.  In Sec.~2, we do not assume a power-law $\rsig$, but instead
solve for $\rsig$ using the Jeans equation with input $\rho(r)$ and
$\beta(r)$ that are motivated by simulations.  We find that while $\rsig$
is approximately a power law within the radial range probed by the current
$N$-body simulations, it is not universally a power-law and shows
significant deviations below this range.  This approach (for isotropic
velocities) was also used in \citet{Graham06}.  While our results for the
$\beta=0$ case agree with theirs in the range of radii explored in their
paper (down to $\sim 0.1$\% of the virial radius $r_{\rm vir}$), we explore
7 orders of magnitude below this range and predict non-power-law behavior
in $\rsig$ in the upcoming simulations.  As we show in Sec.~3, this
conclusion remains unchanged when velocity anisotropy is included in the
calculation, and when a more general form of the density-velocity relation,
$\rho/\sigma_D^\epsilon$, is considered.  For completeness, Sec.~4 compares
the results for the case when $\rsig$ is restricted to be a power law.

\section{Result: $\rsig$ is not a Power-Law}

For a spherical, self-gravitating collisionless system in equilibrium, its
density and velocity structures obey the Jeans equation
\begin{equation}
  \frac{1}{\rho} \frac{(\rho \sigma_r^2)}{dr} + \frac{2\sigma_r^2 \beta}{r}=
    - \frac{d\Phi}{dr} \,,
\label{Jeans}
\end{equation}
where $\rho$ is the radial density profile, $\beta=1-\sigma_t^2/\sigma_r^2$
is the velocity anisotropy parameter, $\sigma_r$ and $\sigma_t$ are the
one-dimensional radial and tangential velocity dispersions, and $\Phi$ is
the gravitational potential.  To apply the Jeans equation, we take as input
a form for $\rho$ and solve for $\sigma_r$.  For the purposes of this
paper, we consider two broad types of radial density profiles.  The first
type is cuspy all the way to the halo center with an inner logarithmic
slope $\gamma$ and an outer slope $\gamma_\infty$:
\begin{equation}
  \rho(r) = \frac{\rho_0}{(r/r_{-2})^{\gamma} [ 1 +  r/r_{-2} ]^{\gamma_\infty-\gamma} } \,,
\label{GNFW}
\end{equation}
where $r_{-2}$, often referred to as the scale radius, is the radius at
which $d\ln\rho/d\ln r = -2$.  Examples of special cases of $(\gamma,
\gamma_\infty)$ that have been proposed for dark matter haloes include
$(-1, -3)$ by \citet{NFW97}, $(-1, -4)$ by \citet{Hernquist90} and
\citet{DC91}, and $(-1.5, -3)$ by \citet{Moore99}.  We find very
similar results from our calculations for $\gamma_\infty=-3$ vs $-4$;
we will thus set $\gamma_\infty=-3$ and refer to equation~(\ref{GNFW}) 
as GNFW below.

The other type of density profile considered in this paper has a non-cuspy
inner profile, given by the Einasto profile \citep{Einasto69} advocated in
several recent studies (e.g., \citealt{Merritt05, Graham06, Navarro08,
  Stadel08}):
\begin{equation}
   \ln \frac{\rho(r)}{\rho_{-2}} = \frac{2}{\alpha} [ 1 - (r/r_{-2})^\alpha ] \,.
\label{Einasto}
\end{equation}
This profile has the feature that its logarithmic slope is itself a
power-law in $r$: $d\ln \rho/d \ln r=-2 (r/r_{-2})^\alpha$.  Unlike
equation~(\ref{GNFW}) that has a definite inner slope of $\gamma$, this
profile continues to flatten towards the halo center.
Equation~(\ref{Einasto}) has the same form as the Sersic profile commonly
used to fit the two-dimensional projected surface brightness of galaxies
\citep{Sersic68}.  The Sersic index $n$ is simply equal to $1/\alpha$.

Starting with either profile in equation~(\ref{GNFW}) or (\ref{Einasto}),
and $\beta=0$ or some form of $\beta(r)$, we integrate the Jeans
equation~(\ref{Jeans}) to obtain $\sigma_r(r)$.  The numerical integration
is performed from large radius (typically $r=10^5 r_{-2}$) down to the
radius $r$ of interest, using a standard Runge-Kutta integrator
\citep{Press92}.  The velocity dispersion is assumed to be zero at the
starting large radius.  The $\sigma_r$ from the integration is then
combined with the input $\rho(r)$ to obtain $\rsig(r)$.

\begin{figure*}
    \subfigure{
    \includegraphics[width=0.45\textwidth]{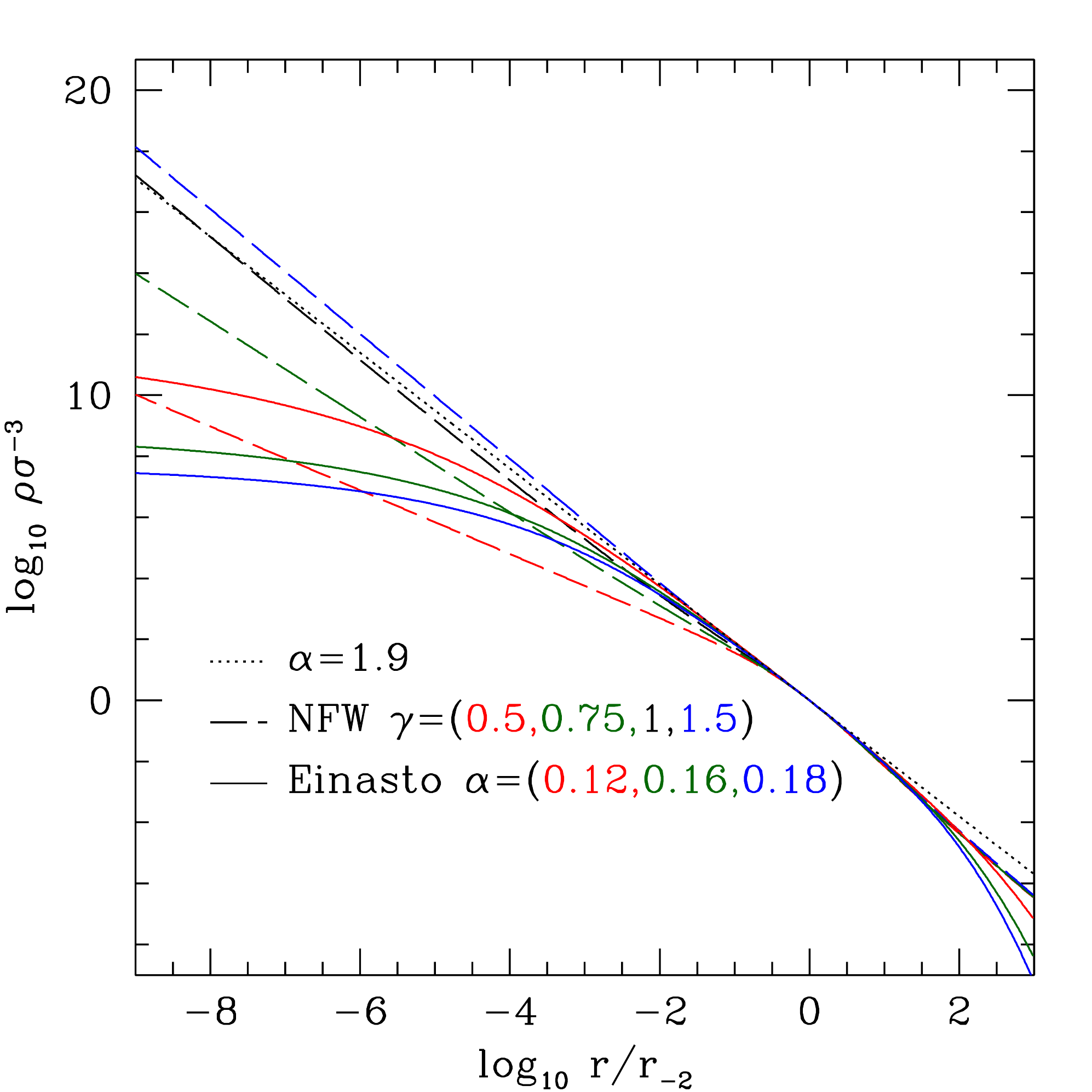} }
    \subfigure{
    \includegraphics[width=0.45\textwidth]{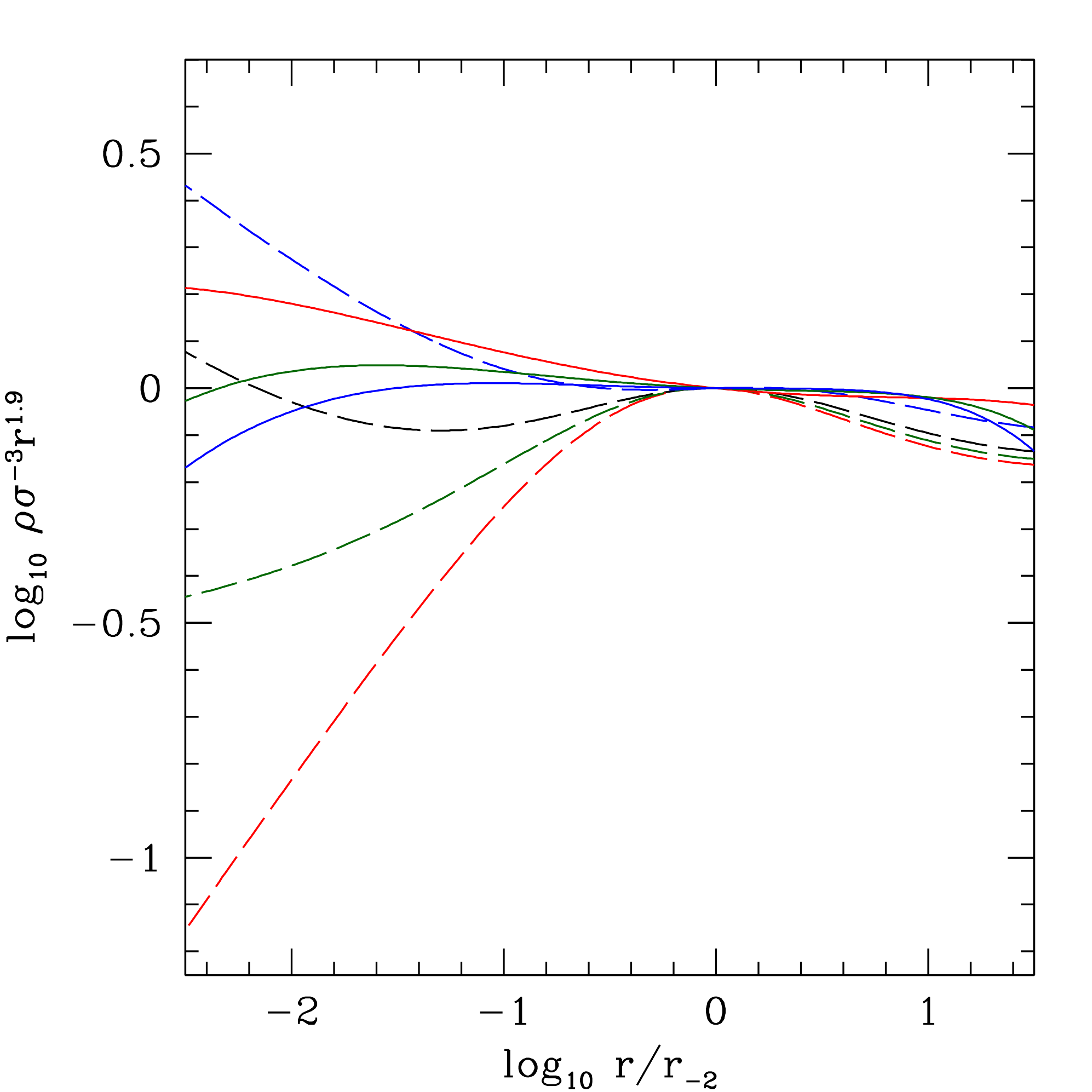} }
    \subfigure{
    \includegraphics[width=0.45\textwidth]{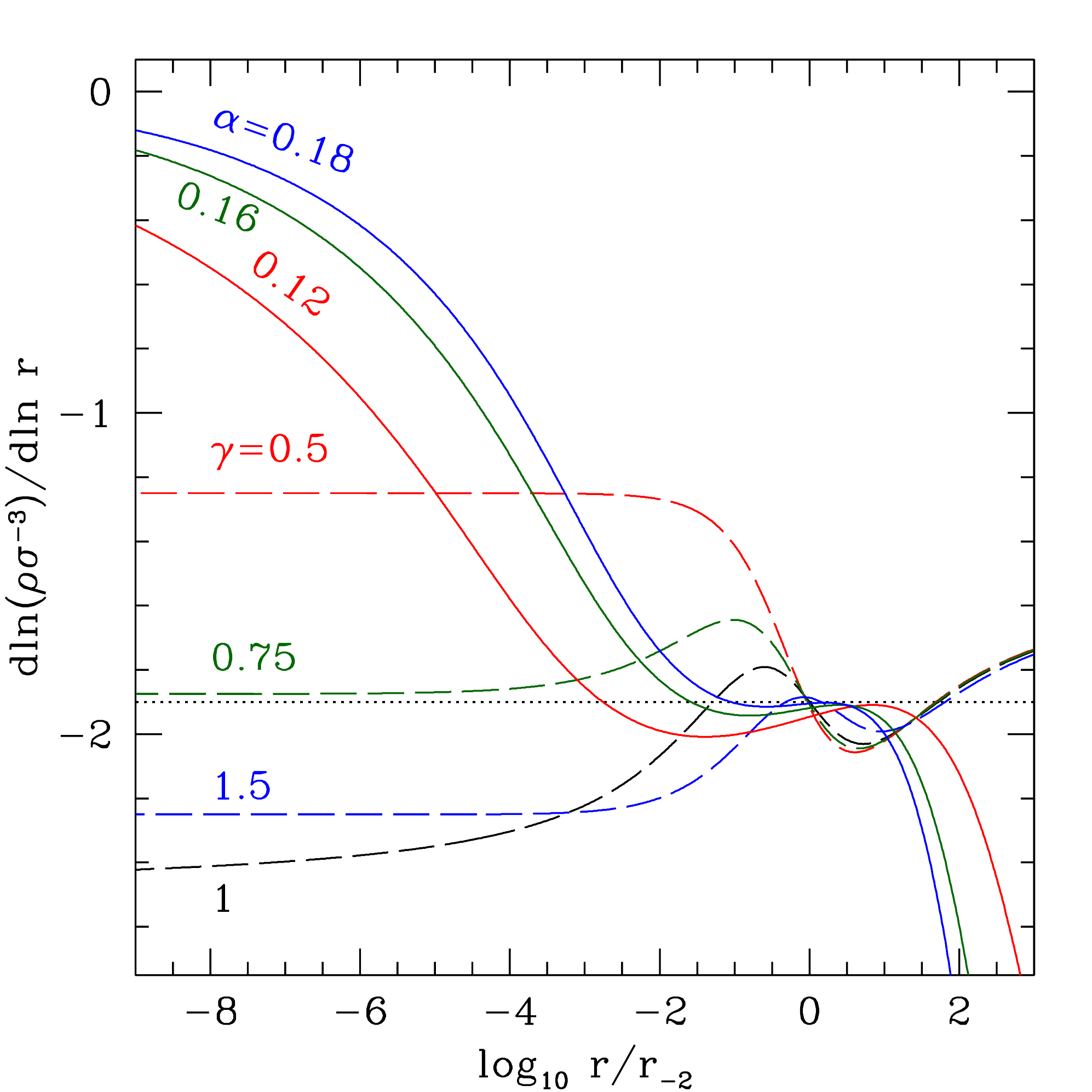} }
    \subfigure{
    \includegraphics[width=0.45\textwidth]{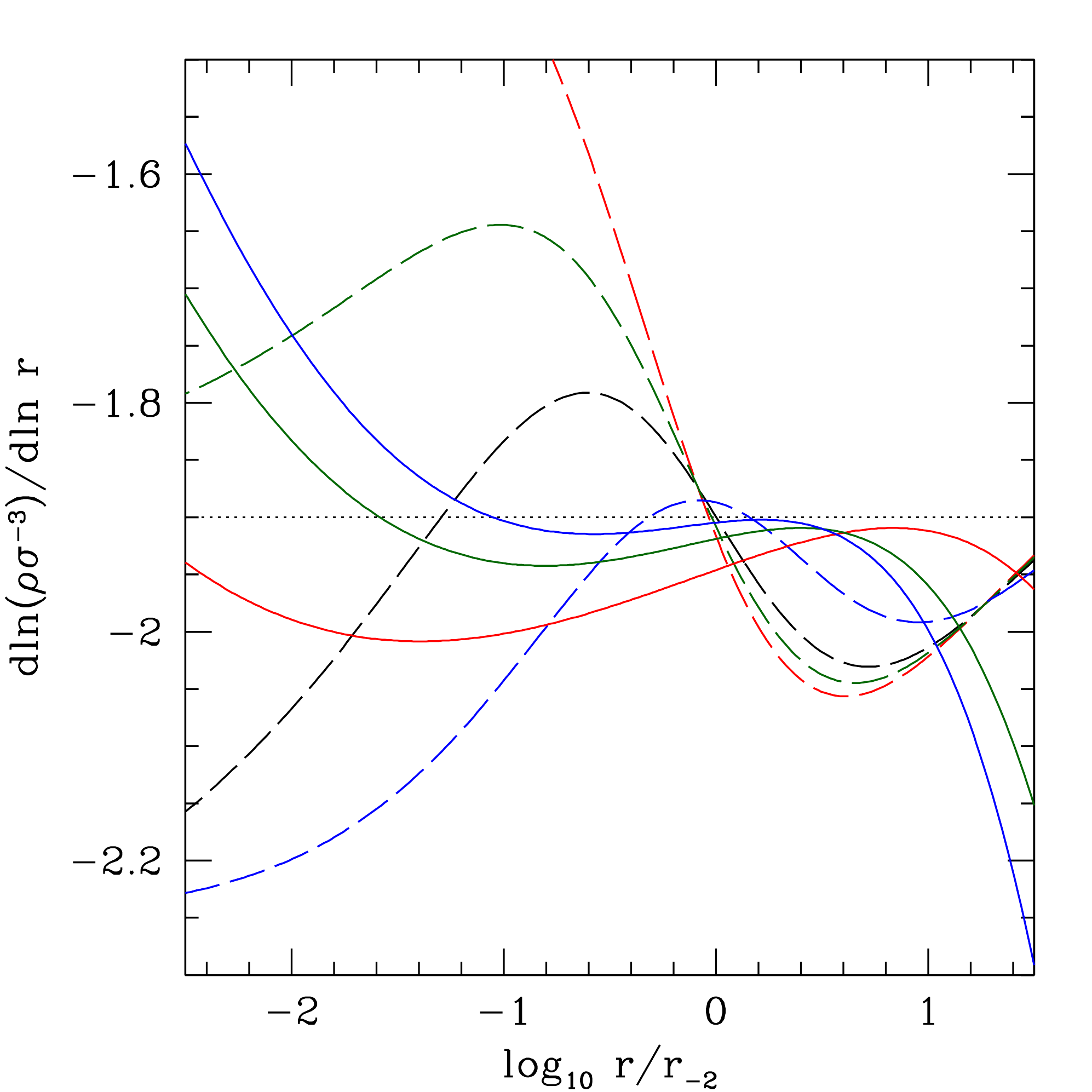} }
%  \includegraphics[scale=0.42]{power_nfw.pdf} 
%  \vspace{-1in}
  \caption{ Radial profiles of the pseudo-phase-space density $\rsig(r)$
    (upper panels) and the corresponding logarithmic slope $d\ln
    \rsig(r)/d\ln r$ (lower panels) obtained from the spherical Jeans
    equation with $\beta=0$ for seven input halo density profiles: Einasto
    (solid) with $\alpha=0.18$ (blue), 0.16 (green), and 0.12 (red), and
    GNFW (dashed) with $\gamma=1.5$ (blue), 1 (black), 0.75 (green), and
    0.5 (red).  The left panels show the behavior of $\rsig(r)$ over 12
    orders of magnitude in $r$, while the right panels show zoom-in views
    of the region $0.01 \la r/r_{-2} \la 10$, which corresponds to the
    range resolvable by the latest $N$-body simulations.  For ease of
    comparison with a power-law, the light dotted straight lines indicate
    the critical case $\rsig \propto r^{-1.9}$, and the $y$-axis in the
    upper right panel plots the logarithm of the ratio of $\rsig(r)$ to
    $\rsig \propto r^{-1.9}$. All curves are scaled to have $\rsig=1$ at
    $r=r_{-2}$.  }
\end{figure*}

The left two panels of Fig.~1 shows the result $\rsig$ (upper panel) and
its logarithmic slope (lower panel) over 12 orders of magnitude in halo
radius for $\beta=0$ and both types of input density profiles: GNFW with
$\gamma=0.5, 0.75, 1.0$, and 1.5 (dashed curves), and Einasto with
$\alpha=0.12, 0.16$, and 0.18 (solid curves).  The range of $\alpha$ is
chosen to span the best-fit values of 0.115 to 0.179 for the six simulated
galaxy-size haloes in \citet{Navarro08}.  The right two panels of Fig.~1
show zoom-in views of the portion of the left figures that is resolvable by
current simulations: $0.01 \la r/r_{-2} \la 10$.

Fig.~1 illustrates that $\rsig$ is not a power-law in $r$ for any of the
seven input density profiles.  For GNFW haloes, the slopes of $\rsig$
exhibit oscillations and deviate noticeably from the critical case $\rsig
\propto r^{-1.9}$ (indicated by light dotted straight lines), in particular
in the extreme cases of inner cusps of $\gamma=0.5$ (red dashed) and $1.5$
(blue dashed).  The Einasto haloes also deviate strongly from a power-law
at small radius.  The zoom-in panels show, however, that the slopes of
$\rsig$ {\it happen} to be quite close to $-1.9$ over the {\it limited}
range of $r/r_{-2} \sim 0.01$ to 10 that is resolvable by current
simulations.  This feature is particularly striking for the Einasto
profiles, where all three solid curves for $\rsig$ have a slope within
$-2.0$ and $-1.8$ from $r/r_{-2}\sim 0.01$ to 10, with the deviations only
starting to show up at the smallest radius $r/r_{-2}\sim 0.01$ near the
simulation resolution limit.  It is therefore not surprising that the
power-law behavior of $\rsig(r)$ continues to appear to be valid even
though the latest simulations find the Einasto form a better fit for
$\rho(r)$ than GNFW -- Fig.~1 shows the Einasto profiles in fact predict a
more power-law $\rsig(r)$ for $r/r_{-2}\ga 0.01$

The important point to note, however, is at the smaller radius of $r/r_{-2}
\la 0.01$ in Fig.~1.  Here $\rsig(r)$ deviates far away from a pure power
law with a wide range of slopes that depend on the input $\rho$.  For
Einasto haloes, the shape of $\rsig$ flattens continuously towards the halo
center, reaching the asymptotic value of $d\ln \rsig/d\ln r=0$ at $r=0$
regardless of the parameter $\alpha$.  This is not unexpected of the
Einasto profile as the density approaches an asymptotic value in the core.
For GNFW, there are two possibilities.  For inner slopes of $\rho$ that are
steeper (shallower) than the critical $\gamma\approx 0.75$, the power-law
slopes of $\rsig$ are steeper (shallower) than the critical $-1.9$.  At the
critical $\gamma \approx 0.75$, the GNFW halo has $\rsig \approx r^{-1.9}$
at small $r$, a result consistent with that of \citet{TN01}, which showed
that starting with an exact power-law of $\rsig \propto r^{-1.875}$, the
resulting density profile has an inner slope of $d\ln\rho/d\ln r \approx
0.75$.  It is worth noting, however, that even the $\gamma=0.75$ GNFW halo
shows wiggles in the corresponding $\rsig$ profile; that is, no GNFW haloes
have an exact power-law $\rsig$.  

%From this analysis, we conclude that while present simulations find density
%profiles that consistently give power-law $\rsig$ at the radii probed,
%future simulations with 10 times or better spatial resolution should reveal
%noticeable deviations from a power law in $\rsig$, if the Einasto density
%profile continues to provide a good fit at small $r$ as suggested by the
%latest simulations.

\section{Further Considerations: $\rsig$ is still not a Power-Law}

We also find the conclusion reached in Sec.~2 to hold not just for
isotropic velocity distributions, but also for anisotropic velocity
distributions. To illustrate this point, we solve equation~(1) using an
input $\beta(r)$ motivated by $N$-body simulations \citep{HM06,Zait08},
where $\beta$ is a function of the local logarithmic slope of the density
profile:
\begin{equation}\label{eq:beta}
  \beta(r) = -0.2 \left( \frac{d\ln \rho}{d\ln r} +  0.8 \right) \,.
\end{equation}
We then compute $\rsig(r)$ for a similar suite of GNFW and Einasto
profiles.  The results are shown in Fig.~2.  What is especially notable is
how insensitive the slopes of $\rsig$ are to the form of $\beta(r)$ used in
the calculation.  This independence from velocity anisotropy strengthens
our conclusion reached earlier, namely, that $\rsig$ is not a power law.

\begin{figure}
  \centering
  \includegraphics[scale=0.42]{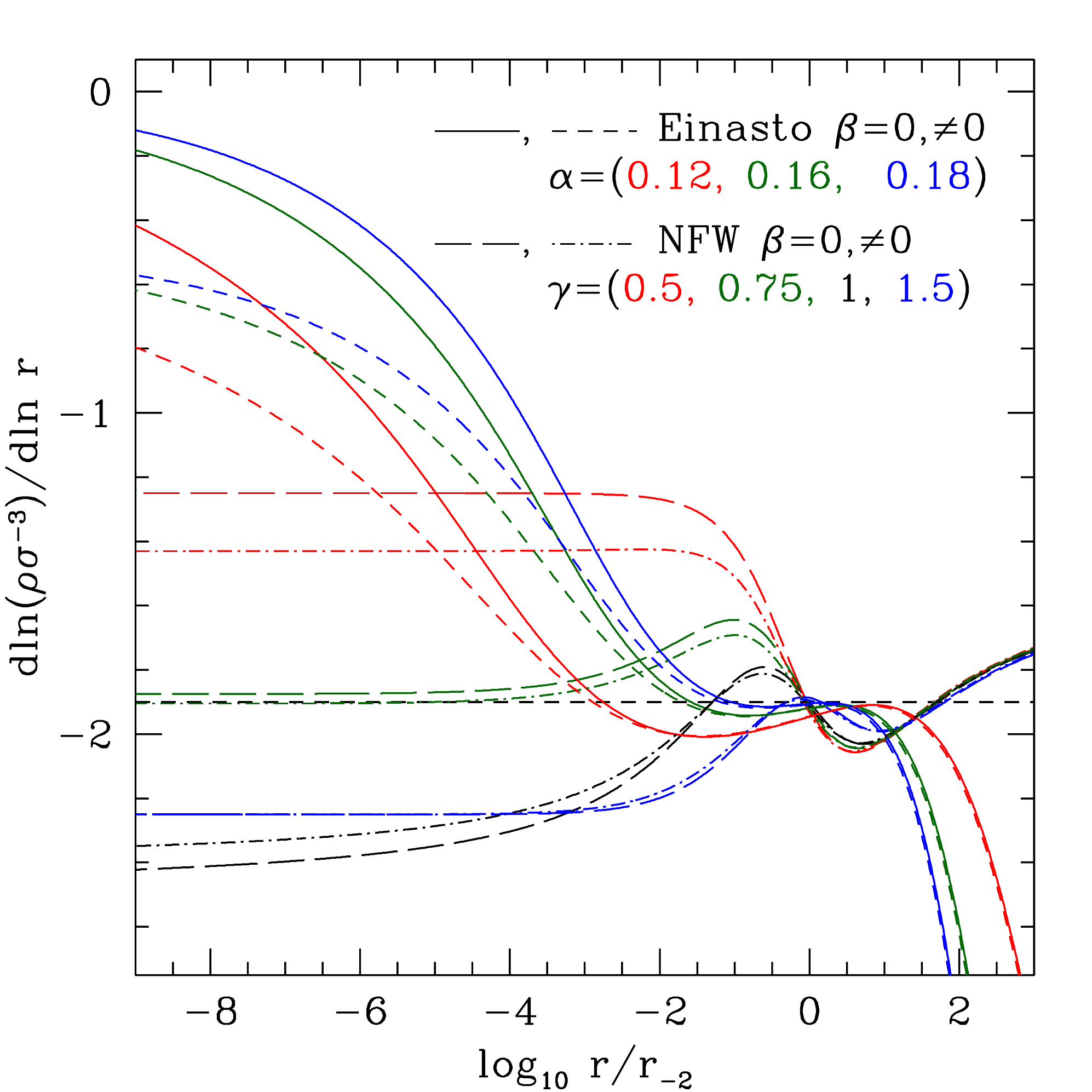} 
  \caption{Effects of velocity anisotropy on the radial profile of $\rsig$
    computed from the Jeans equation: $\beta=0$ (solid for Einasto; long
    dashed for GNFW; same as lower left panel in Fig.~1), and $\beta$ given
    by eq.~(\ref{eq:beta}) (short dashed for Einasto; dashed-dotted for
    GNFW).  The seven input density profiles are the same as in Fig.~1.
    This figure illustrates that including velocity anisotropy in the
    Jeans equation does not change $\rsig$ significantly and still results
    in a non-power-law $\rsig$.}
\end{figure}

Our findings are also in line with some recent work that has called into
question the universality of $\rsig$.  For instance, \cite{Schmidt08}
has advocated that individual simulated haloes are better fit by a
generalized power-law relation that is not necessarily $\rsig$:
\begin{equation}
\frac {\rho}{\sigma_D^{\epsilon}} \propto r^{-\alpha},
\label{rsignew}
\end{equation}
where $\sigma_D = \sigma_r\sqrt{1 + D\beta}$, and $D$ parameterizes a
generalized $\sigma_D$; for instance, $D=0, -2/3$ and $-1$ correspond to
$\sigma_D=\sigma_r, \sigma_{\rm tot}$ (1-d), and $\sigma_t$, respectively.
\cite{Schmidt08} showed that the best-fit values of $(D, \epsilon, \alpha)$
differ from halo to halo, and as a set, they roughly follow the linear
relations $\epsilon=0.97 D + 3.15$ and $\alpha=0.19 D + 1.94$.  For
$\sigma=\sigma_r$ (i.e. $D=0$), the optimal relation is
$\rho/\sigma_r^{3.15} \propto r^{-1.94}$, which is consistent with the
reported behavior of $\rsig$ in $N$-body simulations within error bars.
However, few haloes' best-fit value of $D$ in \cite{Schmidt08} is near
$D=0$.

To assess whether any of these relations is closer to a power-law than
$\rsig$ in our calculations, we choose three sets of $D$ and $\epsilon$
from their linear relation, $(D,\epsilon)= (0, 3.15), (-2/3, 2.50)$, and
$(-1, 2.18)$, and plot in Fig.~3 the logarithmic slopes of these three
relations $\rho/\sigma_D^{\epsilon}$, using our solutions of the Jeans
equation with non-zero $\beta$ shown in Fig.~2.  For clarity, only the
three Einasto profiles are shown in Fig.~3, although our conclusions apply
to the GNFW profiles as well.  The three solid curves in Fig.~3 (for $D=0$)
represent $\rho/\sigma_r^{3.15}$, and are therefore very similar to the
short-dashed curves for the Einasto $\rsig$ in Fig.~2.  The other two sets
of curves in Fig.~3 suggest that $\rho/\sigma^{2.5}_{\rm tot}$
(i.e. $D=-2/3$) and $\rho/\sigma^{2.18}_t$ (i.e. $D=-1$) are also far from
being a power-law over the wide range of radius shown.  When it is limited
to the range of $0.01 \la r/r_{-2} \la 10$ probed by simulations,
$\rho/\sigma^{2.5}_{\rm tot}$ appears to be slightly closer to a power-law
than $\rsig$ for the $\alpha=0.12$ (red) and 0.16 (green) Einasto profiles.
Over the larger range of radius shown in Fig.~3, however, our earlier
conclusion of a non-power-law $\rsig$ carries over to the broader range of
$\rho/\sigma_D^{\epsilon}$ shown.

\begin{figure}
  \centering
  \includegraphics[scale=0.42]{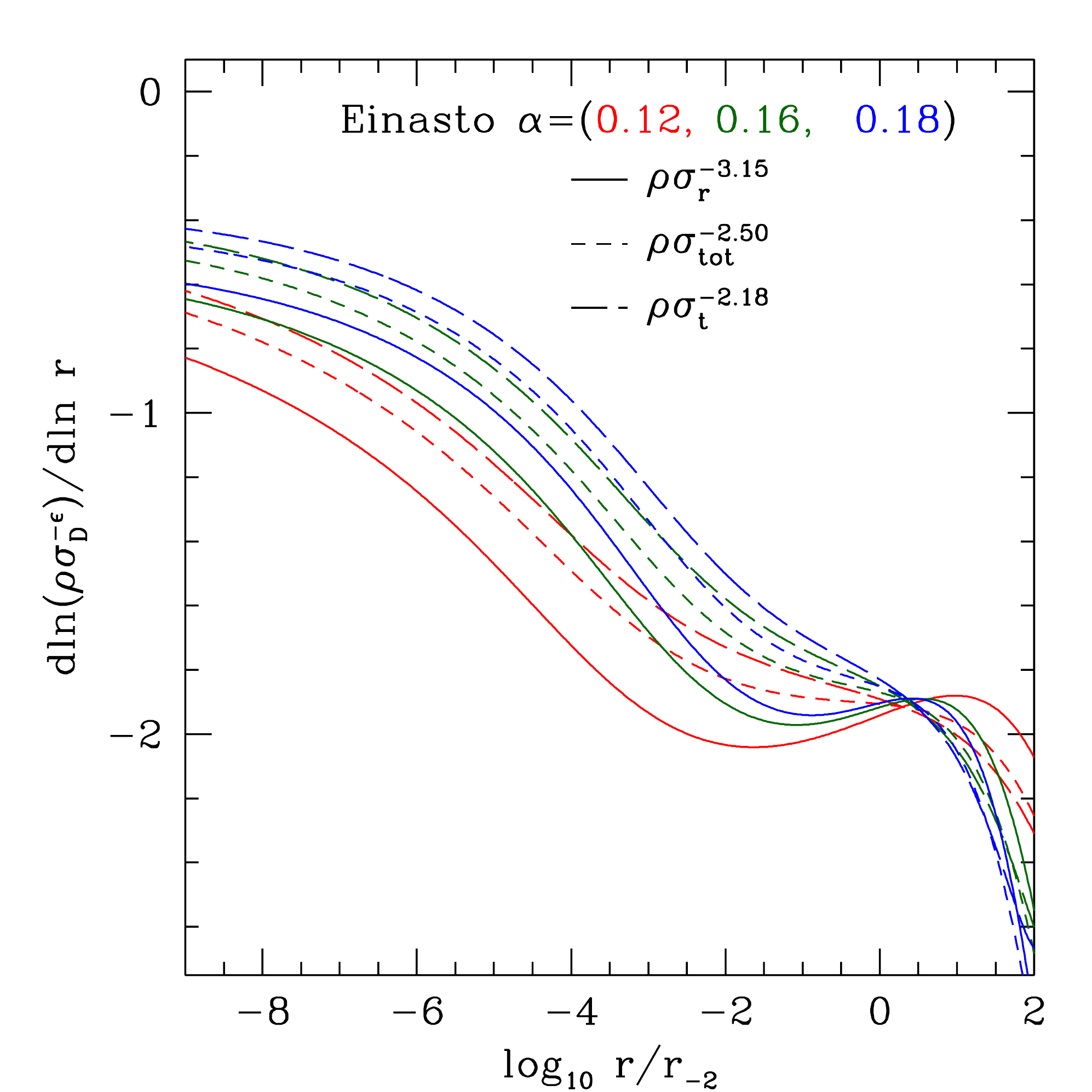} 
  \caption{Radial dependence of three examples of the general
    density-velocity relation $\rho/\sigma_D^\epsilon$ \citep{Schmidt08}
    obtained from our calculations: $\rho/\sigma_r^{3.15}$ (solid),
    $\rho/\sigma_{\rm tot}^{2.5}$ (short-dashed), and
    $\rho/\sigma_t^{2.18}$ (long-dashed).  The $y$-axis shows the
    logarithmic slopes of these relations.  The input $\beta$ is from
    equation~(4), and the input $\rho$ are the three Einasto profiles shown
    in Figs.~1 and 2.  This figure illustrates that like $\rsig$,
    $\rho/\sigma_D^{\epsilon}$ is also far from being a power-law over the
    wide range of radius shown.  }
\end{figure}

\section{Forcing $\rsig$ to be a Power Law}

\begin{figure}
  \centering
  \includegraphics[scale=0.42]{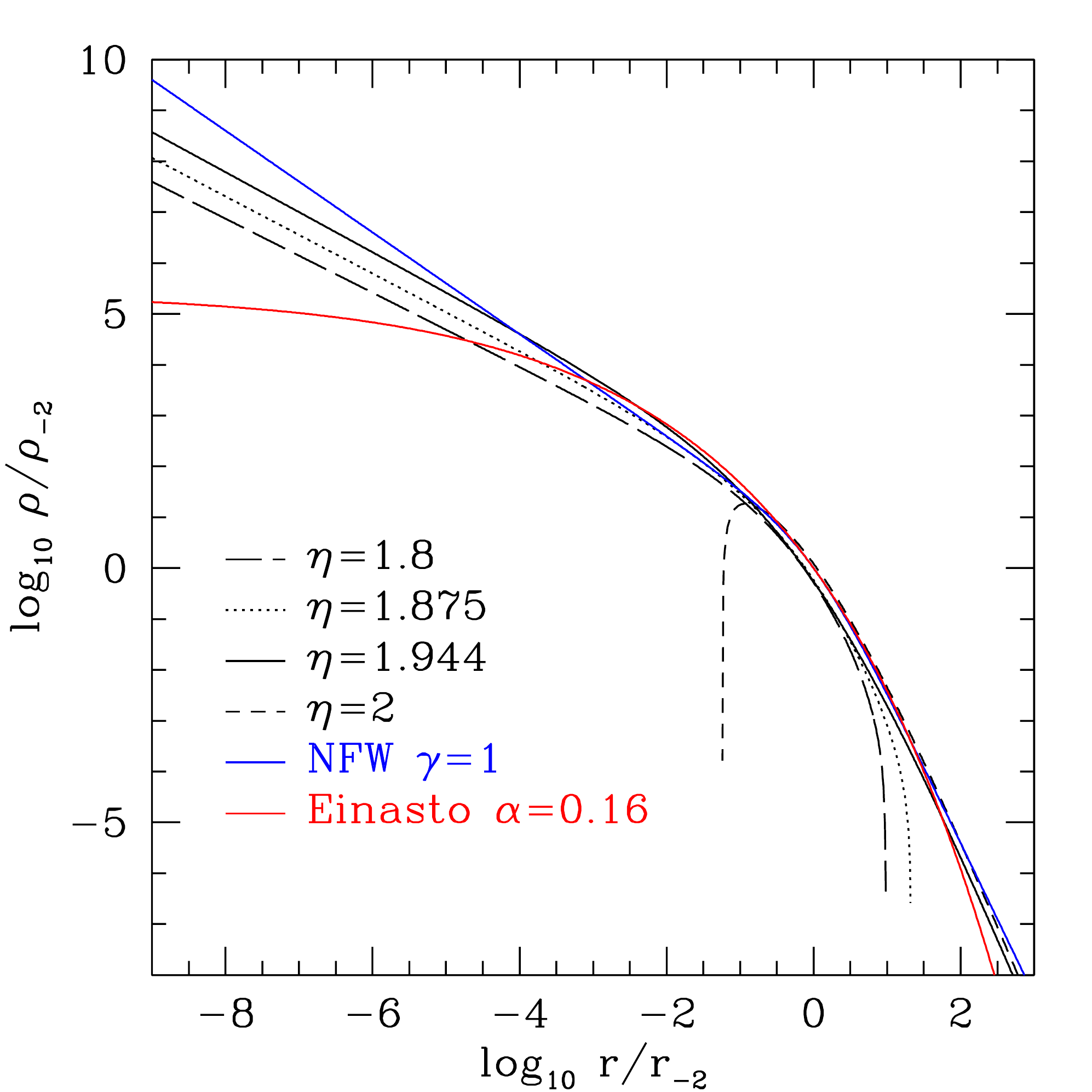} 
  \caption{Solutions for $\rho(r)$ computed from the spherical Jeans
    equation with an input power-law $\rsig \propto r^{-\eta}$ and $\beta=0$.  Four
    values of $\eta$ are shown.
%   where $\eta=1.875$ and 1.9444 are the two
%    critical values examined in \citet{TN01} and \citet{DM05},respectively.  
    Only when $\eta$ is in the range between $\sim 1.8$ and 1.9444 does the
    solution not contain unphysical holes at small $r$ (e.g. $\eta=2$) or
    excessive cutoffs at large $r$.  The shapes of $\rho$ in this narrow
    range of $\eta$, however, deviate significantly from both the Einasto
    (solid red) and $\gamma=1$ GNFW (solid blue) profiles at small radius.
    Dark matter haloes therefore cannot follow the Einasto $\rho$ and a
    power-law $\rsig$ simultaneously at all radii.  }
\end{figure}

Thus far we have solved the Jeans equation assuming an input $\rho(r)$.  As
a point of comparison, we have also solved the Jeans equation following the
works of previous authors with a starting assumption of a power law $\rsig
\propto r^{-\eta}$.  Most notably, \citet{TN01} presented results for the
special case of $\eta=15/8=1.875$, whereas \cite{DM05} (see also
\citealt{Austin05}) argued that only a single {\it realistic} solution
exists for $\rho$, given when $\eta$ takes the particular value of
$35/18=1.9444$ for isotropic velocities, and $\eta=35/18-2\beta_0/9$ for
anisotropic velocities, where $\beta_0 \equiv \beta(r=0)$.

Here we examine a range of $\eta$ and show in Fig.~4 our numerical
solutions (for $\beta=0$) from the Jeans equation for $\rho(r)$ for four
values of input power-law $\rsig \propto r^{-\eta}$: $\eta=1.8, 1.875,
1.9444$ and 2.0.  As $\eta$ moves away from the critical value 1.9444,
$\rho$ drops to zero rapidly at some small radius if $\eta > 1.9444$
(short-dashed curve), while $\rho$ is cutoff sharply at some large $r$ if
$\eta < 1.9444$ (long-dashed and dotted curves).  The former solution with
a central hole is unphysical for dark matter haloes.  The latter, however,
is not automatically ruled out.  Only for $\eta \lesssim 1.8$ do we find
the outer $\rho$ to drop off too steeply to represent realistic
$\Lambda$CDM haloes.  It therefore appears that the narrow range of $1.8
\la \eta \le 1.9444$ may admit physical solutions for $\rho$.  It is
important to keep in mind, however, that $\rho$ in these cases are not well
described at small radius by either the Einasto profile advocated by recent
simulations, or the original $\gamma=1$ GNFW profile.  Lowering the inner
slope of the GNFW profile to $\gamma \approx 0.75$ provides a better fit.
Dark matter haloes therefore cannot be well fit by the Einasto $\rho$ and a
power-law $\rsig$ simultaneously at all radii.

\section{Conclusions}

Motivated by the apparent power-law radial profile of the pseudo-phase
space density, $\rsig$, reported in various N-body simulations (e.g.,
\citealt{TN01, Ascasibar04, DM05, Hoffman07, Stadel08,Navarro08}), we solve
the Jeans equation to study the {\it implied} pseudo-phase-space density
for given parameterizations of $\rho$ suggested by $N$-body simulations,
i.e., Einasto and GNFW.  We find that independent of the velocity
anisotropy, $\rsig$ is not a pure power law in radius for either the
Einasto or GNFW profiles (left panels of Fig.~1).  In the radial ranges
that are probed by current $N$-body simulations (down to $\sim 10^{-3}
r_{vir} \sim 10^{-2} r_{-2}$), we find that $\rsig$ happens to be
approximately a power law, in particular for the Einasto profile (right
panels of Fig.~1).  For radial scales right below the resolution of current
simulations, however, we see significant deviations from a power law
profile for $\rsig$ if either Einasto or GNFW continues to hold as a
suitable parameterization of $\rho$.  This result is unchanged when
velocity anisotropy is included in the calculation (Fig.~2), and when a
more general density-velocity relation of $\rho/\sigma_D^{\epsilon}$ is
considered (Fig.~3).  The hope to gain deep insight into the process of
dark matter halo formation using a simple and universal power law scaling
of $\rsig$ may therefore be misleading.  Conversely, when $\rsig \propto
r^{-\eta}$ is assumed as an input in the Jeans equation, none of the
realistic solutions for $\rho$, which occur only for the narrow range of
$1.8 \la \eta \le 1.9444$, take the Einasto form (Fig.~4).

We therefore conclude that the two halo properties seen in current
simulations -- the Einasto $\rho$ and power-law $\rsig$ -- cannot hold
simultaneously at all radii.  We expect that the upcoming simulations with
better spatial resolution will settle this debate: either the Einasto
profile will continue to hold and $\rsig$ will show a break from a power
law, or $\rsig$ will continue as a power law inward and $\rho$ will deviate
from its current parameterizations.

%\section*{Acknowledgments}

We thank Onsi Fakhouri, Peng Oh, and Simon White for useful
discussions.  PC and JZ are supported by the Theoretical Astrophysics
Center at UC Berkeley.

\bibliographystyle{mn2e} 
\bibliography{rhosig}

\label{lastpage}

\end{document}